\documentstyle[11pt,newpasp,epsf]{article}

\begin{document}

\title{The Infrared Spectrum of R Doradus}

\author{Nils Ryde}
\affil{McDonald Observatory and Department of Astronomy,
University of Texas at Austin, USA}

\author{Kjell Eriksson}
\affil{Uppsala Astronomical Observatory, Sweden}







\begin{abstract}

Here, we present our modelling (Ryde \& Eriksson, 2002) of the
$2.6-3.7\,\micron$ spectrum of the red semiregular variable R
Doradus observed with the Short-Wavelength Spectrometer on board
the Infrared Space Observatory. We will also present the entire
spectrum of R Dor up to $5\,\micron$ based on our model
photosphere in order to show which molecules are important for the
emergent spectrum.
\end{abstract}




%
%
%
\index{*R Dor}

\section{Observations and analysis}

Our goal is to discern whether a moderately varying M giant, such
as R Doradus, could be modelled satisfactory  with a hydrostatic
model photosphere. We observed the  $2.6 - 3.7\,\micron$ region
with the ISO SWS 06 at a resolution of $R = 2000 - 2500$ and the
full $2.4 - 45\,\micron$ region with the SWS 01 at $R = 30$.  The
low-resolution SWS01 observation is used to ensure that the
different SWS06 observations in this region were properly aligned.

We have calculated hydrostatic model photospheres with the {\sc
marcs} code in spherical geometry. Synthetic spectra were
calculated using these model photospheres. The synthetic spectra,
also calculated in spherical geometry, include lines from H$_2$O
(Partridge \& Schwenke 1997), CO (Goorwitch 1994), SiO (Langhoff
\& Bauschlicher 1993), CH (J\o rgensen et al. 1996), CN (J\o
rgensen \& Larsson 1990 and Plez 1998, priv. com.), OH (Goldman et
al. 1998), C$_2$ (Querci et al. 1971 and J\o rgensen 2001, priv.
com.), and CO$_2$ (Hitemp; Rothman et al. 1992).

\begin{figure}
\plotone{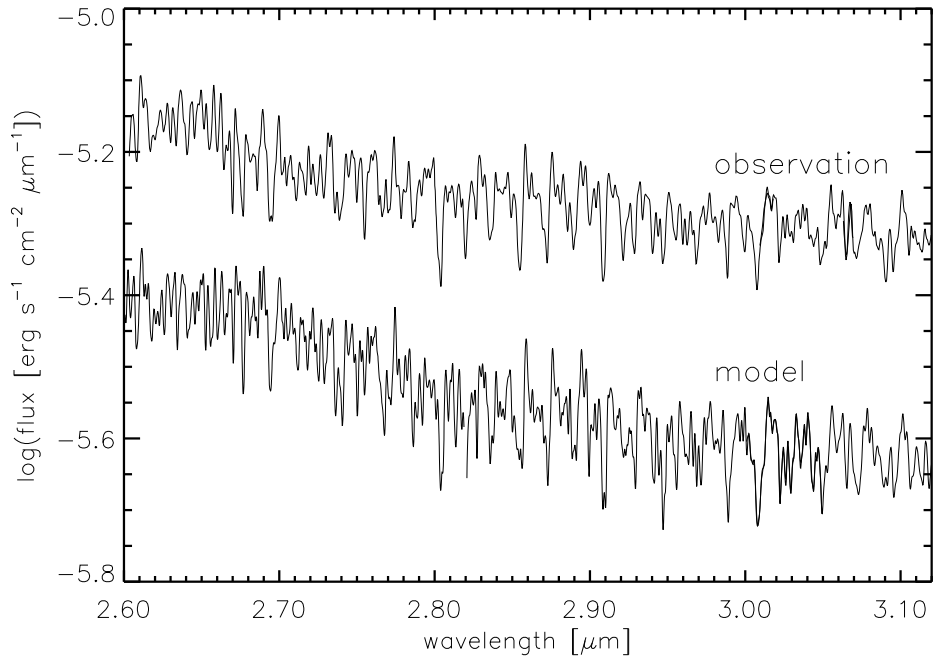} \caption{A comparison between the observed
and the synthetic spectra of R Dor.} \label{fig-noll}
\end{figure}

\begin{figure}
\plotone{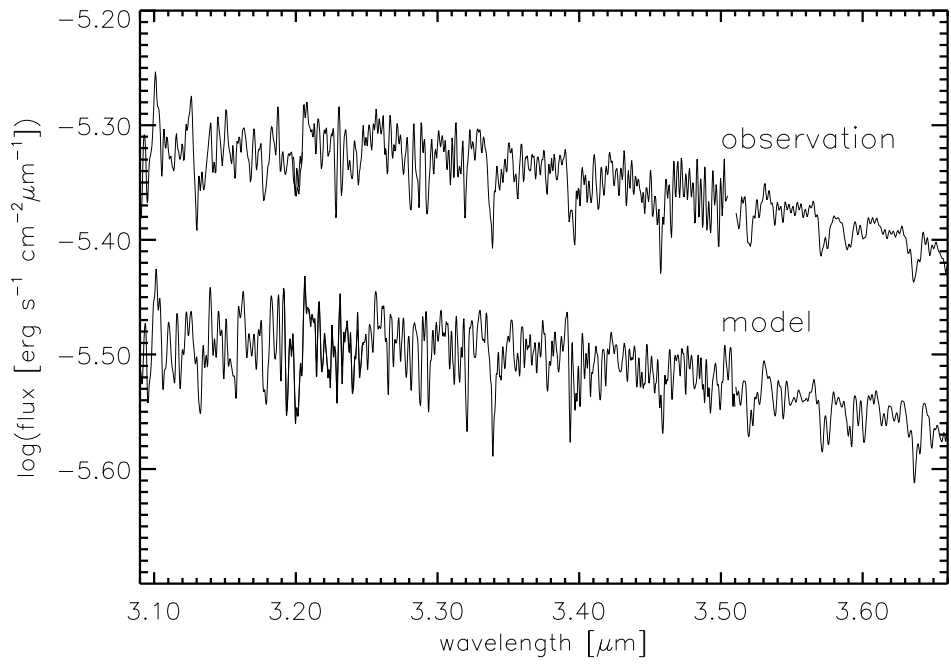} \caption{A comparison between the observed
and the synthetic spectra of R Dor.} \label{fig-0}
\end{figure}

We are able to fit the observed range well with a hydrostatic
model photosphere, see Figs. \ref{fig-noll} and \ref{fig-0}. Our
best fit model have the fundamental parameters of $T_{eff} = 3000$
K, $\log(g)=0$, and solar metallicity. We find that the wavelength
region investigated is sensitive to the effective temperature used
in the modeling of the photosphere. The agreement between the
synthetic spectrum and the ISO observations is encouraging,
especially in the wavelength region of $2.8-3.7\,\micron$,
suggesting that a hydrostatic model photosphere is adequate for
the calculation of synthetic spectra in the near infrared for this
moderately varying red giant star. However, an additional
absorption component is needed at $2.6-2.8\, \micron$. Thus, we
also find that the model does not account for this extra
absorption component at the beginning of the interval. For a
discussion of this discrepancy see Ryde \& Eriksson (2002). The
spectral signatures are dominated by water vapour in the stellar
photosphere, but several photospheric OH, CO, and SiO features are
also present.

\begin{figure}
\plotone{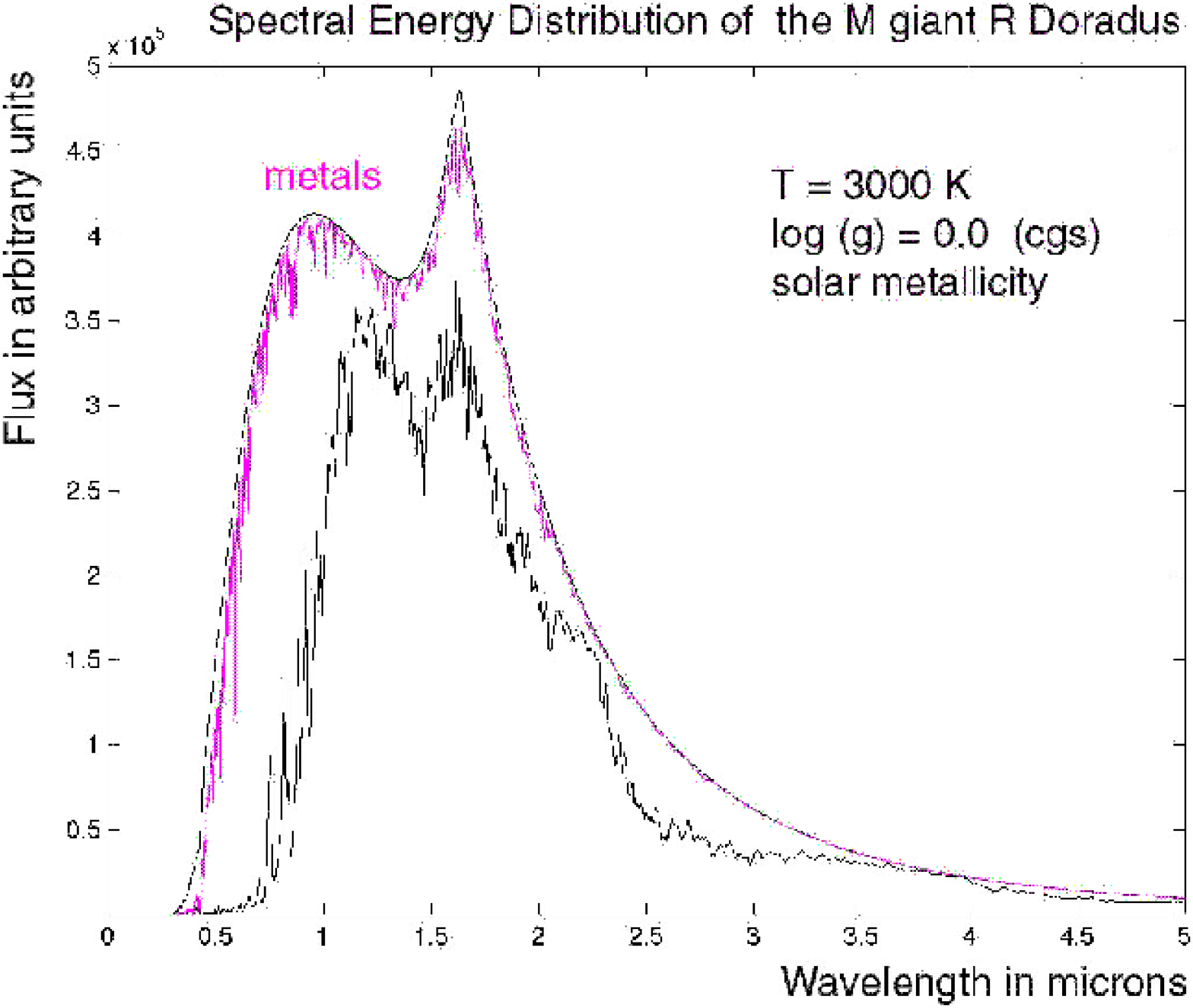} \caption{R Dor's spectrum up to 5 microns
is shown by the black, lower line. In red is shown the synthetic
spectrum calculated by only taking metals into account. The full
upper line is the synthetic spectrum with only the continuous
opacities taken into account.}\label{fig-1}
\end{figure}

\begin{figure}
\plotone{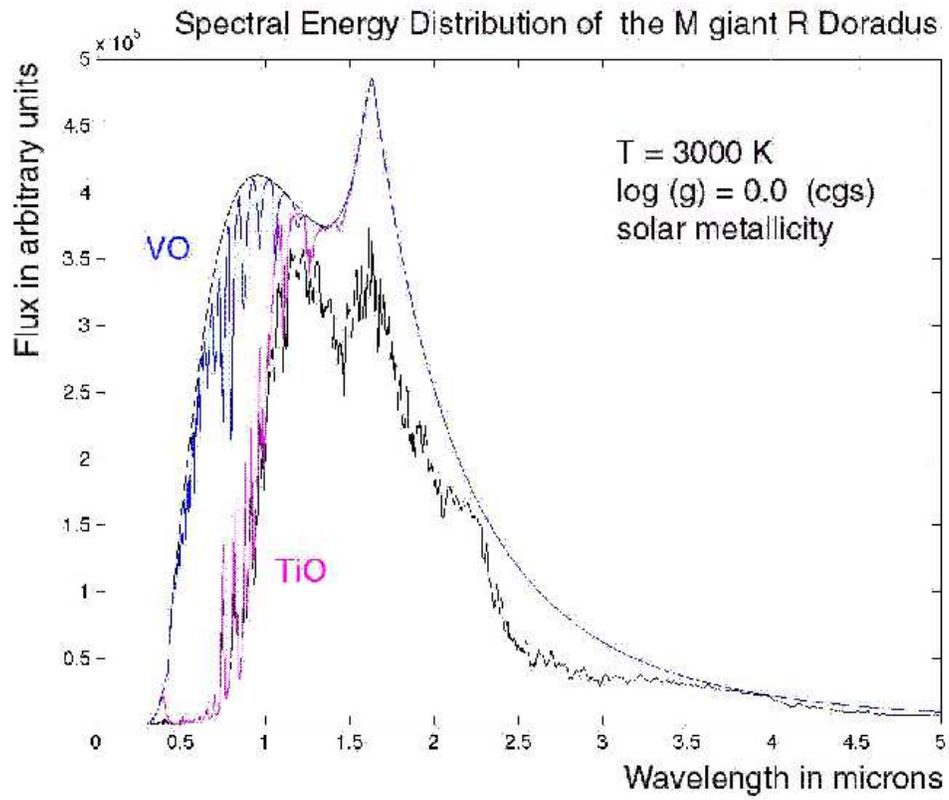} \caption{Same as in Fig. \ref{fig-1} but
showing TiO and VO.} \label{fig-2}
\end{figure}

\begin{figure}
\plotone{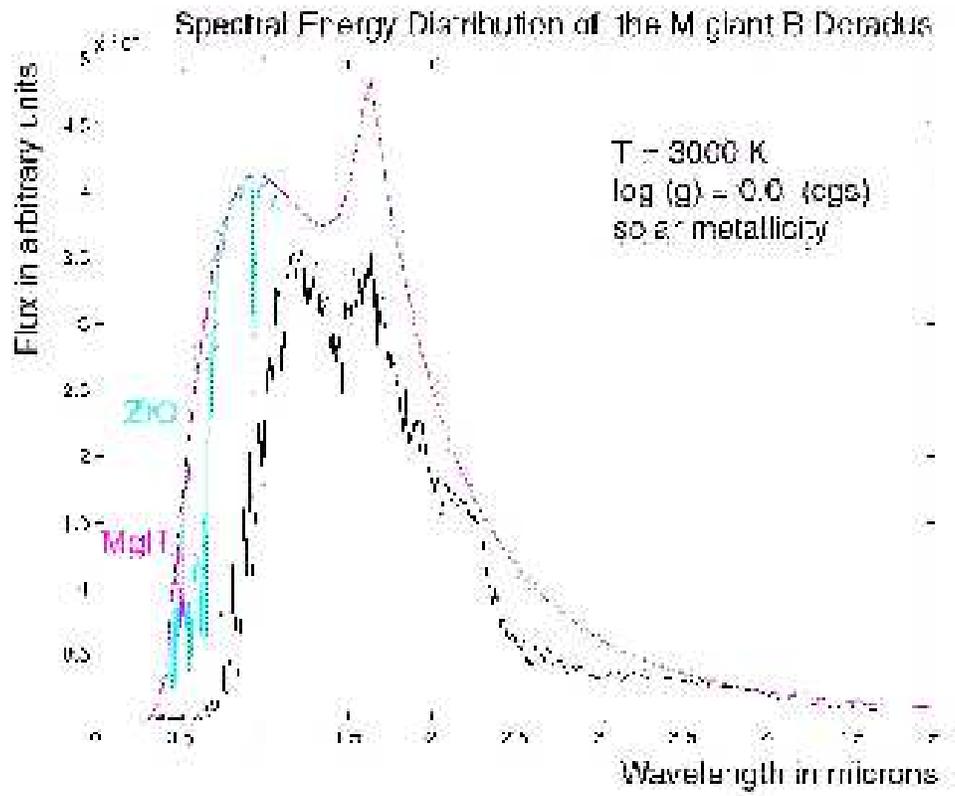} \caption{Same as in Fig. \ref{fig-1} but
showing ZrO and MgH.} \label{fig-3}
\end{figure}

\begin{figure}
\plotone{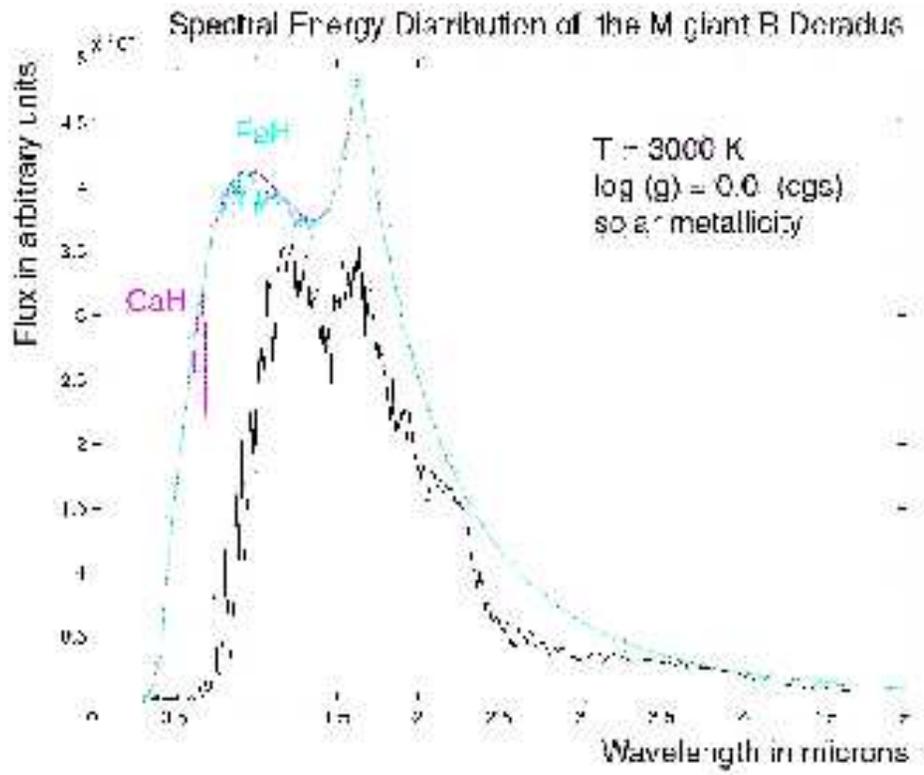} \caption{Same as in Fig. \ref{fig-1} but
showing FeH and CaH} \label{fig-4}
\end{figure}

\begin{figure}
\plotone{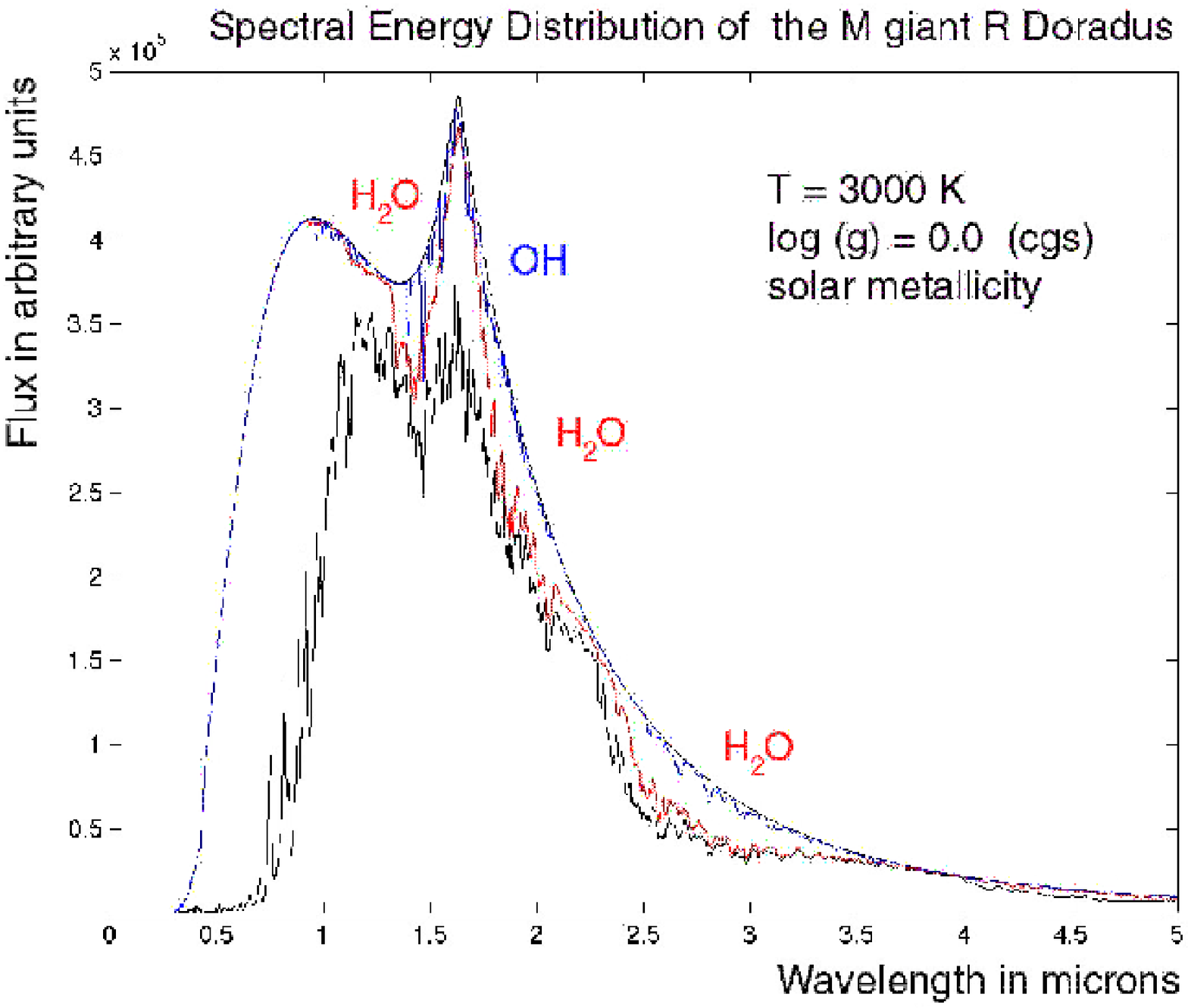} \caption{Same as in Fig. \ref{fig-1} but
showing OH and H2O.} \label{fig-5}
\end{figure}

\begin{figure}
\plotone{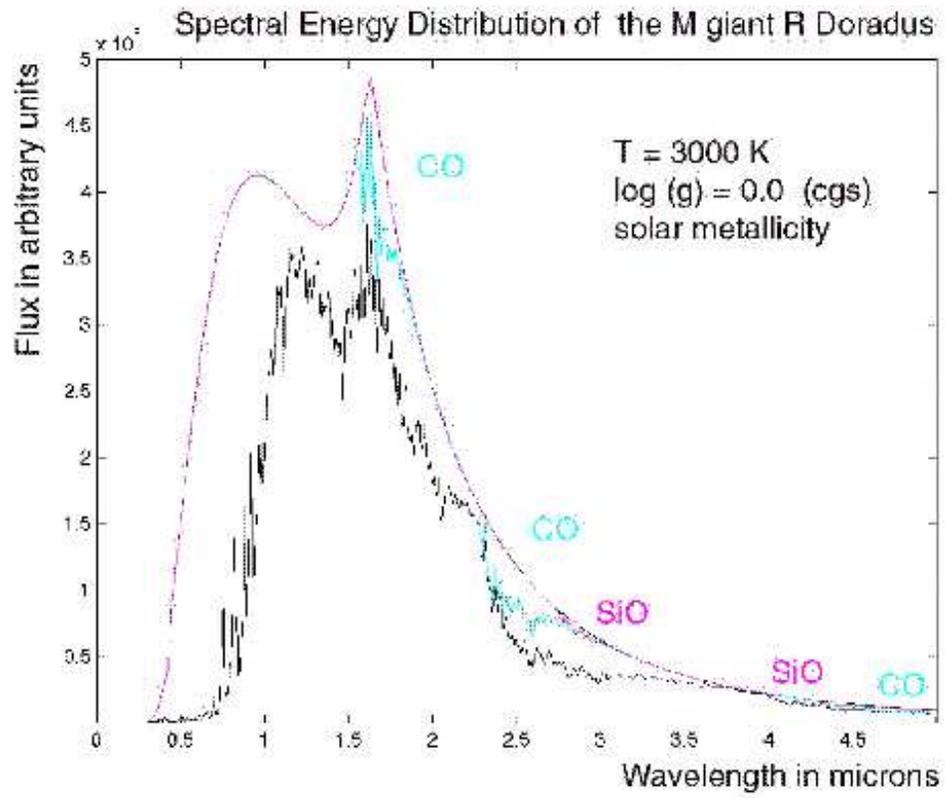} \caption{Same as in Fig. \ref{fig-1} but
showing CO and SiO.} \label{fig-6}
\end{figure}

\begin{figure}
\plotone{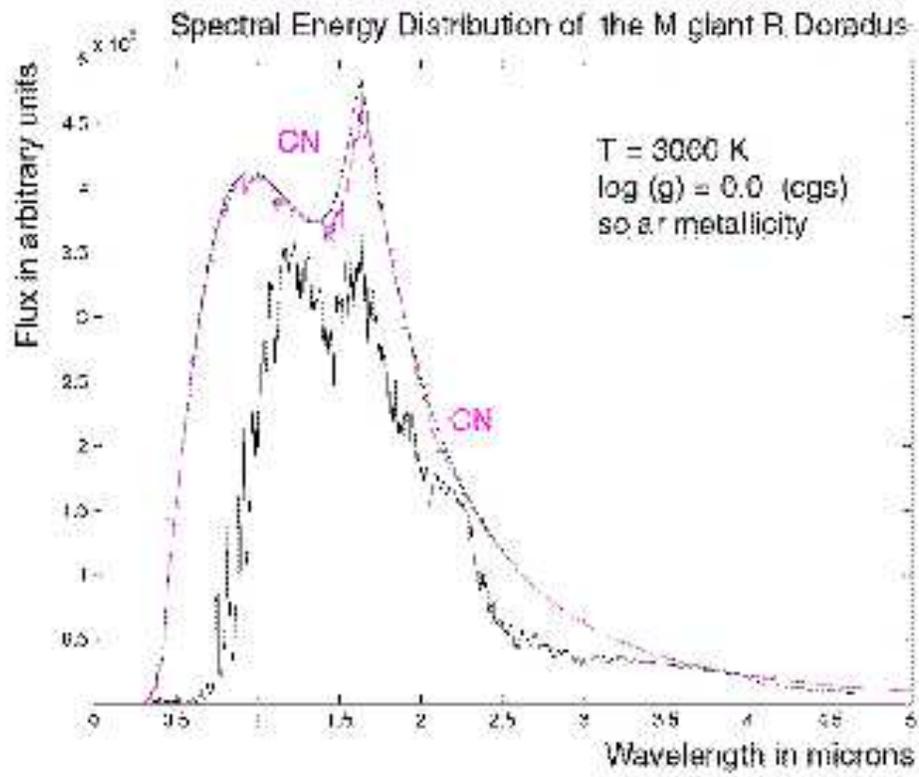} \caption{Same as in Fig. \ref{fig-1} but
showing CN.} \label{fig-7}
\end{figure}

Figs. $\ref{fig-1}-\ref{fig-7}$ show which molecules are important
contributors to the emergent spectrum (up to $5\,\micron$) of R
Doradus. The synthetic spectra shown there are all calculated
using our R Dor model atmosphere. However, the spectra are
synthesized with only one molecule at a time.


\section{Concluding remarks}
R Dor is a moderately varying M Giant. We have modelled its 3
micron ISO spectrum (resolving power is ca 2000) successfully with
a hydrostatic model photosphere. Dynamical models are, however,
certainly needed for the Mira stars, which show larger variations
with time. Line opacities contributing at $2.6 - 3.7$ microns are
H$_2$O, CO, OH, and SiO. The region is found to be very
temperature sensitive, and it is found that an extra absorption is
needed at 2.6 - 2.8 microns.

\newpage
\acknowledgments

We should like to thank Drs. B. Gustafsson and D. L. Lambert for
inspiration and enlightening discussions. This work was supported
by the P.E. Lindahl Foundation Fund of the Royal Swedish Academy
of Sciences
and the Swedish Foundation for International Cooperation in
Research and Higher Education.

%
%

%

\end{document}